\documentclass[grl]{AGUTeX}

\usepackage{lineno}
\usepackage{graphicx}
\usepackage{xcolor}

\authorrunninghead{GASTINE ET AL.}

\titlerunninghead{EXPLAINING JUPITER'S MAGNETIC FIELD}

\begin{document}

\title{Explaining Jupiter's magnetic field and equatorial jet dynamics}

\authors{T. Gastine,\altaffilmark{1}
J. Wicht,\altaffilmark{1}  L.~D.~V. Duarte,\altaffilmark{1,2}
M. Heimpel\altaffilmark{3}, and A. Becker\altaffilmark{4}}

\altaffiltext{1}{Max-Planck-Institut f\"{u}r
Sonnensystemforschung, Justus-von-Liebig-Weg 3, 37077 G\"{o}ttingen, Germany.}

\altaffiltext{2}{Laboratoire de G\'eologie de Lyon, CNRS,
Universit\'e de Lyon, France.}

\altaffiltext{3}{Department of Physics, University of Alberta, Edmonton, Alberta
T6G 2J1, Canada.}

\altaffiltext{4}{Institut f\"ur Physik, Universit\"at Rostock, 18051 Rostock,
Germany.}


\begin{abstract}
Spacecraft data reveal a very Earth-like Jovian magnetic
field \citep{Connerney98,Connerney07}.
This is surprising since numerical
simulations have shown that the vastly different interiors of terrestrial
and gas planets can strongly affect the internal dynamo process.
Here we present the first numerical dynamo that manages to match the
structure and strength of the observed magnetic field
by embracing the newest models for Jupiter's
interior \citep{Nettelmann12,French12}.
Simulated dynamo action  primarily occurs in the deep
high electrical conductivity region while
zonal flows are dynamically constrained to a strong equatorial jet
in the outer envelope of low conductivity.
Our model reproduces the structure and strength of the observed
global magnetic field and predicts that secondary dynamo action
associated to the equatorial jet produces banded magnetic features
likely observable by the Juno mission.
Secular variation in our model scales to about $2000\,$nT per year and should
also be observable during the one year nominal mission duration.
\end{abstract}

\begin{article}

\section{Introduction}

Spacecraft data allow to model the Jovian magnetic field up to
spherical harmonic degree and order $4-5$
\citep{Connerney98,Connerney07,Hess11}.
The observations reveal that the Jovian magnetic field is
dominated by a tilted dipole with an Earth-like inclination angle around
$10^\circ$ and is about an order of magnitude stronger than
the geomagnetic field.
Tracking cloud features with ground-based and space
observations show that the Jovian surface dynamics is dominated by strong
zonal motions of unknown and highly debated depth. These zonal winds form a
differential
rotation profile with alternating prograde (i.e. eastward) and retrograde
(westward) flows
\citep[e.g.][]{Porco03}.

Jupiter's atmosphere mainly consists of hydrogen and helium
(roughly a quarter by mass), surrounding a rocky inner core with
a radius likely less than 10\% of Jupiter's mean radius $R_J$ (1 bar
level) \citep{Nettelmann12}. Laboratory experiments and ab initio
simulations have shown that hydrogen undergoes a phase transition from a
molecular to a metallic state at increasing pressure and temperature
\citep{Nellis95,French12}. The Jovian dynamo is thought
to operate below a transition radius, located between
$0.85$ and $0.90\,R_J$, while the observed fierce zonal winds likely
dominate the dynamic of the outer molecular layer. Increasing density,
Lorentz forces and Ohmic dissipation would lead to slower fluid velocities in
the metallic layer and confine the zonal winds to the upper region
\citep[e.g.][]{Liu08}. Classical dynamical models therefore treat the dynamics
of these two layers separately, focussing either on the dynamo action
\citep{Christensen06,Gastine12a} or on the zonal winds driving
\citep{Heimpel05,Jones09,Kaspi09,Gastine12,Gastine14}.
However, since electrical conductivity and density increase continuously
with depth throughout the transition \citep{Lorenzen11,French12}, as
illustrated in Fig.~\ref{fig:profiles}, the coupling between both layers may
actually be significant.
\cite{Stanley09} incorporate both the density and the electrical conductivity
variations but only simulate the dynamics in the very outer envelope where the
conductivity decays exponentially. They report multiple zonal jets but
the magnetic field is too axisymmetric, too little dipolar, and too
concentrated at higher latitudes.
The model by \cite{Heimpel11} spans the metallic and the molecular envelope but
ignores the density variation. They only find dipole dominated solutions
close to the onset of dynamo action where the field structure and dynamics
is too simplistic.
The approach by \cite{Duarte13} is very similar to the model presented here
but the density profile follows an ideal equation of state and the
electrical conductivity only drops by two orders of magnitude over the
simulated shell. Once more, dipole-dominated solutions are restricted to
low Rayleigh numbers where the overall dynamics remains too simple.

Here, we present a numerical simulation that models the density contrast
up to 99\% of the Jupiter's radius ($1\,$bar level) and uses a more realistic
electrical conductivity profile that mimics the severe drop over
the molecular layer. The good agreement of the large scale field with current
magnetic
field models indicates that the model provides useful constraints for the Juno
spacecraft, which is scheduled to arrive at Jupiter in august 2016.

\section{Dynamo Model}

\begin{figure*}
\centering
\includegraphics[width=10cm]{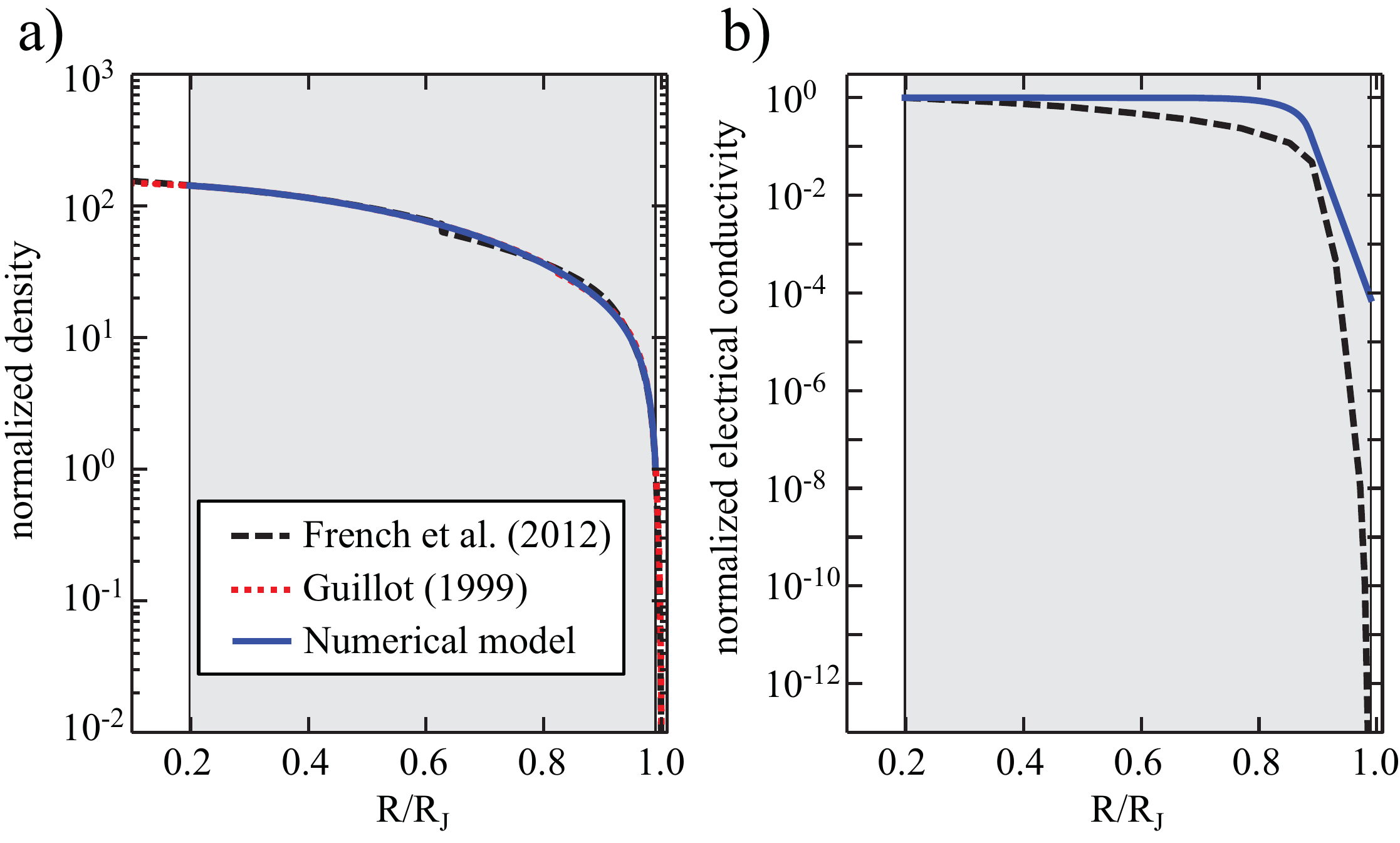}
 \caption{\scriptsize Comparison of the density (panel a)  and
electrical conductivity profile (panel b) used in the numerical
simulation with interior models for Jupiter \citep{Guillot99,French12}.
The density has been normalized with the reference value at $0.99\,R_J$,
while the electrical conductivity has been normalized
with the reference value at $0.2\,R_J$. These radii represent the
outer and inner boundary in the numerical model (blue line).}
 \label{fig:profiles}
\end{figure*}

We consider convection and dynamo action in a spherical shell rotating
at a constant rate $\Omega$ about the $z$-axis. The numerical MHD code
MagIC uses pseudospectral methods and a mixed implicit/explicit
time-stepping scheme to solve for the Navier-Stokes equation, the induction
equation, and the entropy equation \citep{Christensen07}.
We adopt an anelastic approach \citep{lantz99,Jones11} to include a
background density profile $\tilde{\rho}$ that represents
a seventh order polynomial fit to the ab initio predictions \citep{French12}.
The background temperature $\tilde{T}$ is then approximated by a polytropic
equation of state of the form $\tilde{T}=\tilde{\rho}^{1/m}$, where $1/m=0.45$.
Modelling Jupiter's large density contrast illustrated in
Fig.~\ref{fig:profiles}a is computationally challenging. By using a fine
spatial grid we could afford to model the dynamics up to 99\% of Jupiter's
radius, closely following the predicted profile. The inner 19.8\% in radius are
occupied by a solid electrically conducting inner core.
Adopting a conducting inner core eases the
comparison with previous numerical calculations \citep{Duarte13}.
The actual size and thermodynamic properties of Jupiter's core are still a
matter of debate \citep{Nettelmann12,Cebulla14}. However, since the inner core
in our numerical model is small the effects of its conductivity are likely negligible
\citep{Schubert2001,Wicht02}.
The density increases by a factor of 137 over the simulated shell, which
corresponds to 4.92 density scale heights. The density drops by an additional
factor of 174 over the last outer 1\% in radius. Resolving this severe gradient
numerically would be extremely costly.

The electrical conductivity profile predicted by ab initio simulations
(Fig.~\ref{fig:profiles}b) indicates a continuous transition
from the molecular to the metallic state with a clear change in slope
where the hydrogen molecules become completely dissociated. While the
conductivity increases only mildly with radius at depth, the gradient quickly
becomes super-exponential beyond $0.9\,R_J$. Resolving this steep decrease poses
additional numerical problems. As a compromise, we assume that the
conductivity remains constant in the deeper interior and
decreases exponentially by four orders of magnitude beyond
the transition radius $0.87\,r_o$ where $r_o$ is the outer radius in the
simulation \citep{Gomez10,Duarte13}. This choice is motivated by the
magnetic Reynolds number profile further discussed below.

The dimensionless anelastic equations solved by MagIC are governed by four
dimensionless numbers: the Ekman number $E=\nu/\Omega d^2$; the modified
Rayleigh number $Ra^*=T_o \Delta s/\Omega^2 d^2$; the Prandtl number
$Pr=\nu/\kappa$;
and the magnetic Prandtl number $Pm=\nu/\lambda(r_i)$. These parameters
combine characteristic physical properties of the system: the rotation rate
$\Omega$, the shell thickness $d$, the temperature at the outer boundary
$T_o$, the entropy contrast over the shell $\Delta s$, the kinematic
viscosity $\nu$, the thermal diffusivity $\kappa$, and the magnetic
diffusivity at the inner boundary $\lambda(r_i)$.
Stress-free mechanical boundary conditions are used at the outer
boundary while rigid conditions model the interface to the
conducting inner core. The magnetic field matches a diffusive solution at the
inner boundary and a potential field at the outer boundary. Convection is
driven with an imposed constant entropy contrast over the shell. Although
internal heating would be more realistic for Jupiter's heating mode, 
some preliminary numerical simulations with internal heating at $E=10^{-4}$
yield very similar solutions as the cases without. This suggests that the shape
of the convective pattern is primarily controlled by the local
buoyancy variations caused by the density contrast while the distribution of
buoyancy sources plays a secondary role. However, further parameter
studies will be needed to systematically check the influence of internal
heating on rotating compressible convection.

%
Starting with an Ekman number of $E=10^{-4}$ and a magnetic Prandtl number of
$Pm=2$ we lowered these values gradually down to $E=10^{-5}$ and
$Pm=0.6$. The final setup was integrated for $0.3$ magnetic diffusion times.
The Prandtl number has been fixed to $Pr=1$ and the Rayleigh number
of $Ra^*=6.16\times 10^{-3}$ has been chosen to obtain strongly-driven
convective motions and a complex enough yet still dipole-dominated magnetic
field \citep{Duarte13}.

The spatial resolution of this numerical model is
defined by the maximum spherical harmonic degree ($\ell_{max}=426$) and the
number of radial levels ($N_r=145$).
%

\begin{figure*}
\centering
\includegraphics[width=12cm]{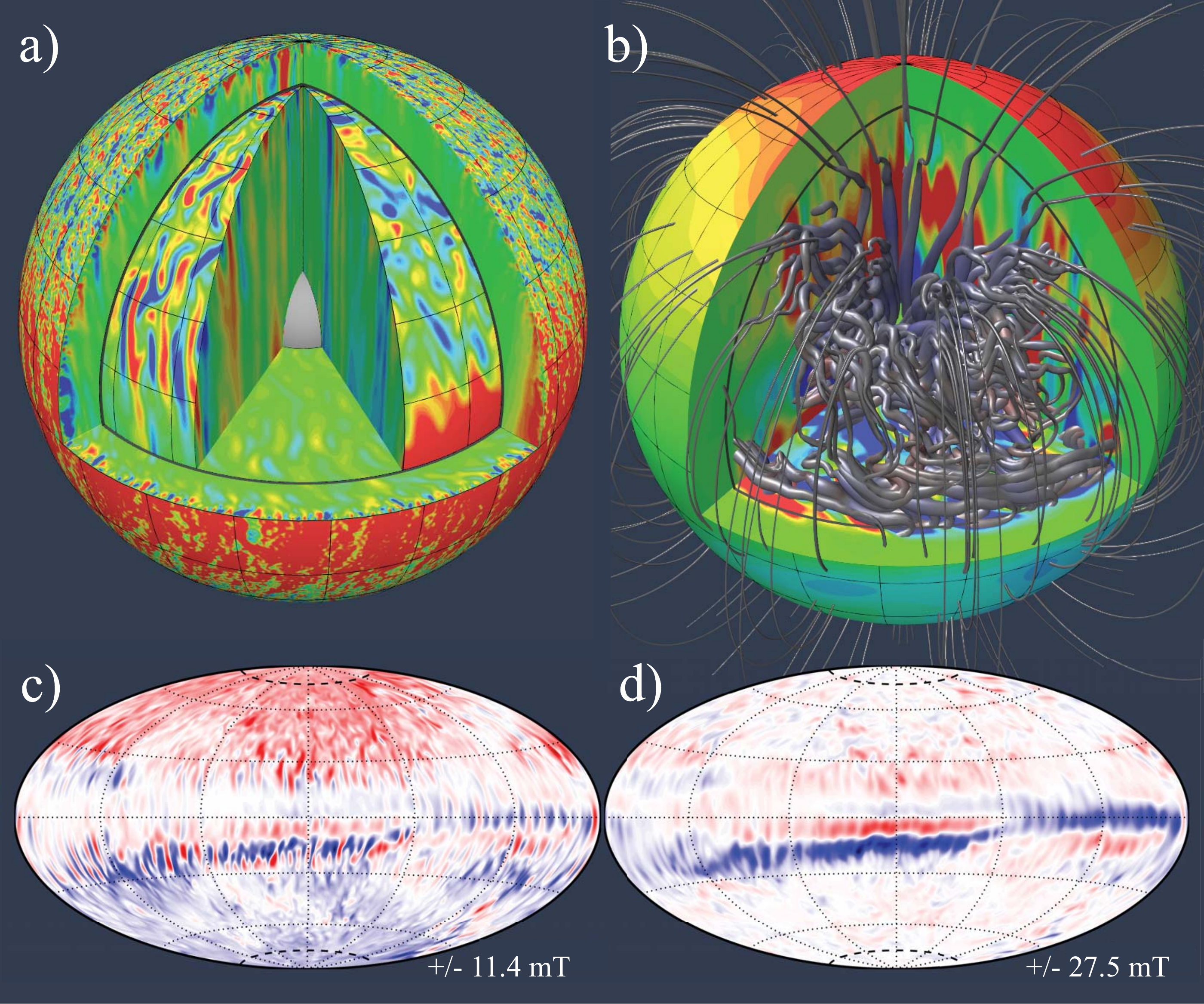}
 \caption{\scriptsize Panel a) shows the azimuthal flow component on the outer
surface and the right cut while the radial flow component is shown in
the equatorial and left cuts.
The inset sphere slices visualize the weaker flow at greater depths. The
right inset shows the azimuthal flow amplified by a factor 10 and the
left inset  shows the radial flow amplified by a factor 2.5.
The flow amplitude strongly increases with radius while the length
scale decreases.
Panel b) shows the radial magnetic field on the outer surface and the
left cut. The surface field has been amplified by a factor 10.
The right and horizontal cuts (at $-10^\circ$) show
the azimuthal magnetic field. The thickness of the magnetic fieldlines has been
scaled with the third root of the local magnetic field strength.
Panels c) and d) show the radial and azimuthal magnetic field at the transition
radius $0.87\,r_o$ that is marked with a dark grey line in panels a) and b).
Yellow/red (blue) indicates outward (inward) or
eastward (westward) directions.}
 \label{fig:snapshot}
\end{figure*}

\section{Rescaling to Physical Units}

Due to numerical limitations it remains impossible to choose realistic values
for all the physical properties in a direct numerical simulation. 
The Ekman
number is $E=10^{-5}$ and the magnetic Prandtl number $Pm=0.6$ in our dynamo
model where values around
$E=10^{-18}-10^{-19}$ and $Pm=10^{-7}$ would be appropriate for Jupiter.
Boussinesq and anelastic dynamo models computed in the accessible
range $E\in[10^{-6}-10^{-3}]$ suggested that the rms field magnetic field
strength and rms convective flow amplitude in the simulated shell depend
on the available power. The respective scaling laws predict reasonable values
for planets or fully convective stars \citep{Christensen06,Christensen10}
and can be used to extrapolate the numerical results.
Our model obeys the anelastic scaling laws \citep{Yadav13a} for the  dimensionless rms flow amplitude
\begin{equation}
\label{Urms}
 \mbox{U}^\star = 1.65\,{P^\star}^{0.42},
\end{equation}
where $P^\star=P /(\Omega^3 d^2)$ is the dimensionless form of the convective
power per unit mass $P$. The dimensionless rms magnetic field strength
B$^\star$ has been found to scale like
\begin{equation}
\label{Brms}
 \mbox{B}^\star = \sqrt{f_{Ohm}\, \rho^\star}\;{P^\star}^{0.35}\;\;.
\end{equation}
The factor $f_{Ohm}$ is the ratio of Ohmic to viscous dissipation and
is close to one for planetary dynamo regions where the magnetic diffusivity
is orders of magnitude larger than the viscous diffusivity. $\rho^\star$ is the
mean dimensionless density.

The available convective power can be estimated based on the net heat flux
density $F$ out of a planetary dynamo region \citep{Christensen10}.
Using the Jovian net heat flux $F_J=5.5\,$W/m$^2$ \citep[e.g.][]{Hanel81},
$M_J=1.83\times10^{27}\,$kg and $R_J=6.99\times10^4\,$km and the radial
profiles by \cite{French12} for the heat capacity $c_p$, gravity $g$ and
thermal expansivity $\alpha$ yields

\begin{equation}
 P = \frac{4\pi R_J^2 F_J}{M_J} \int \frac{\alpha g}{c_p} dr \approx
 4.6\times 10^{-10}\,\mbox{W/kg}\;\;.
\end{equation}
Eq.~(\ref{Urms}) then provides an estimate for the dimensionless
rms flow speed which can be converted into a dimensional value using
Jupiter's rotation rate $\Omega=1.75\times
10^{-4}\,$s$^{-1}$ and shell thickness $d=(0.99-0.2)R_J=5.52\times10^{4}\,$km:
\begin{equation}
\mbox{U}_J = \mbox{U}_J^\star \Omega d \approx 3\,\mbox{cm/s}\;\;.
\end{equation}
Following the same procedure but using Eq.~(\ref{Brms}), the mean Jovian
density $\rho_J=1300\,$kg/m$^3$ and the magnetic permeability $\mu$ predicts
Jupiter's rms magnetic field strength:
\begin{equation}
\mbox{B}_J = \mbox{B}_J^\star \Omega d \left(\rho_J \mu\right)^{1/2}
\approx 7\,\mbox{mT}\;\;.
\end{equation}
Both values reasonably agree with estimates for Jupiter which seems to confirm
the applicability of these scaling laws \citep{Christensen06}.

Since our simulation closely follows the scaling laws we can simply rescale
the result by assuming that the dimensionless rms flow and magnetic field
amplitudes are identical to U$_J$ and B$_J$ respectively.
The flow velocity also allows to estimate Jupiter's typical convective
time scale $t_c= d / \mbox{U}=53$~yr, often referred to as the turnover time.
To predict the time variability of the Jovian magnetic field we
have assumed that the convective overturn time in our simulation agrees
with the respective Jovian value. This scaling implies that the simulation
covers roughly $6500\,$yr.

\section{Results and Implications for Jupiter}

Fig.~\ref{fig:snapshot} illustrates the flow and magnetic field generation in
our numerical model. Owing to the strong density stratification, the length
scale of the non-zonal convective flow decreases with radius while the
amplitude increases \citep[see equatorial cut in
Fig.~\ref{fig:snapshot}a and][]{Gastine12}.
These convective motions maintain a prograde equatorial zonal
flow via Reynolds stresses \citep[see right cut in
Fig.~\ref{fig:snapshot}a and][]{Heimpel05}.
Lorentz forces constrain the equatorial jet to the weakly conducting
outer envelope and largely suppress flanking higher latitude
jets \citep{Duarte13}.

The magnetic Reynolds number $Rm=\mbox{U} d \mu \sigma$ measures the ratio of
magnetic field production to Ohmic dissipation and is thus an important
quantity for any dynamo process. Here, U is a typical flow velocity,
$\mu$ the magnetic permeability, $\sigma$ the electrical conductivity,
and $d$ the depth of the simulated spherical shell.
Numerical simulations suggest that $Rm$ must typically exceed $50$ to guarantee
dynamo action \citep{Christensen06}. In the model presented here, $Rm$ assumes
a complex radial profile since U increases with radius while $\sigma$
decreases. $Rm$ reaches a peak value around $220$ in the inner conducting
layer, decreases to $Rm=50$ near $r=0.9\,r_o$, and reaches $Rm=0.2$ at the outer
boundary. The combination of the mean flow velocity $U$ and the electrical
conductivity profile thus guarantees no dynamo action in the very outer envelope
but leaves room for induction effects in the transitional zone.

Fig.~\ref{fig:snapshot} indeed demonstrates that the primary dipole-dominated
magnetic field
is created at depth where $Rm$ is significant, the electrical conductivity
high, and the density contrast relatively mild.
A secondary dynamo mechanism operates at low latitudes and slightly below the
transition radius where $Rm$ is still sufficient for dynamo action.
Here, the remaining zonal shear (right inner sphere in
Fig.~\ref{fig:snapshot}a) creates a strong azimuthal magnetic field.
At the transition region, the magnetic
banding (Fig.~\ref{fig:snapshot}d) associated with the shear of the
equatorial jet is reminiscent to the so-called ``wreaths of magnetism'' found
in the solar dynamo models by \cite{Brown10}. The non-zonal convective motions
convert this azimuthal field into banded radial
field structures (see thick horizontal fieldlines in Fig.~\ref{fig:snapshot}b
and radial magnetic field at the transition radius in
Fig.~\ref{fig:snapshot}c).

The inhibition of dynamo action in the very outer envelope is essential
since previous numerical models have shown that the vigorous
small scale flows and the strong equatorial jet located in this layer would
promote multipolar magnetic fields \citep{Gastine12a,Duarte13}.
In our model,
the decrease in electrical conductivity is not as steep as
suggested by the ab initio simulations but still sufficient to compensate
for the radial increase in flow amplitude.
It is also steeper than in previous simulations \citep{Duarte13} which
allowed us to adopt a larger Rayleigh number. While this promotes a more vigorous and dynamic
convection, the lower conductivity in the very outer shell guarantees that
this region plays no role in the dynamo process. This not only promotes a more
vigorous flow but also a more complex and time dependent solution that
is much closer to Jupiter's magnetic field. At lower Rayleigh numbers for
example the axial dipole component is much too dominant, as in the dynamo
models by \citet{Heimpel11}.

Estimates for Jupiter predict a much larger magnetic Reynolds number
($7\times10^5$ using Eq.~\ref{Urms}) than in our model.
Increasing the mean $Rm$ in the simulation would require a yet steeper
electrical conductivity profile to minimize dynamo action
in the molecular layer. Additionally, smaller viscosities may have to
be adopted to retain dipole-dominated dynamo
action \citep{Christensen06,Duarte13}.
Though suggested by the ab initio simulations, both of these
measures are currently too expensive for numerical resources.

Jupiter's surface dynamics is dominated by a system of banded strong zonal
winds where the dominant equatorial jet is flanked by several secondary
jets at higher latitudes.
These secondary jets are not captured by our simulation.
Previous studies suggest that a smaller viscosity help to
promote the secondary jets in numerical simulations \citep{Jones09,Gastine14}.
However, the dynamo action of deeper reaching zonal winds may lead to a magnetic
field geometry that is not very Jupiter-like \citep{Gastine12a,Duarte13,Stanley09}.
The strong azimuthal magnetic fields likely produced by deeper reaching zonal winds would
also cause a thermal Ohmic dissipation signal which has not been detected \citep{Liu08,Liu13}.
However, Ohmic heating is not expected to be strong for the dominant equatorial jet
which remains confined to the weakly conducting outer envelope.
Some authors predict that the secondary jets remain
restricted to Jupiter's weather layer not represented in our
model \citep{Kaspi09}. The gravity signal measured by the Juno mission
will provide additional constraints on the depth of the Jovian zonal winds
\citep{Kaspi13,Liu13}.

\begin{figure}
\centering
\includegraphics[width=6cm]{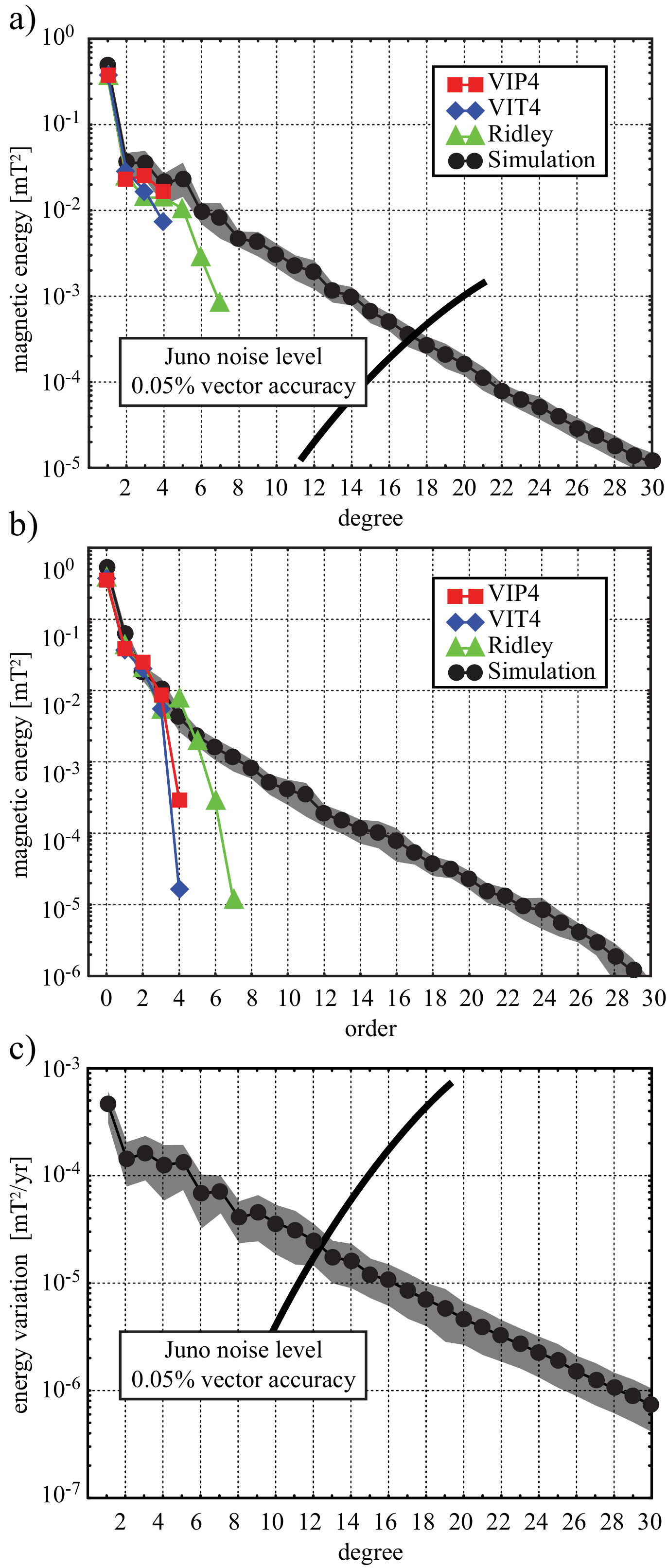}
 \caption{\scriptsize Panel a) and b) show a comparison of the
magnetic energy spectra per spherical harmonic degree $\ell$ and order $m$
for the three different Jupiter field models VIP4 \citep{Connerney98},
VIT4 \citep{Connerney07} and Ridley \citep{Ridley12} with
the time averaged spectrum of the numerical simulation.
The thick black lines in panel a) illustrate the predicted noise level
for the Juno magnetometer. Panel c) shows the time averaged absolute variation
in the $\ell$ energy spectrum per year for the numerical simulation.
The grey bar around the simulated spectra has the width of the
standard deviation and provides an idea of the time dependence.
The thick curved black lines in panels a) and c) illustrate an
estimate for the Juno noise level assuming that the magnetometer delivers
the conservatively expected $0.05$\% vector accuracy. The secular variation
noise level will be somewhat higher than indicated in panel c) since this
estimate assumes that all the data acquired during the nominal one year mission
duration are used.}
 \label{fig:spectrum}
\end{figure}

\begin{figure}
\centering
\includegraphics[width=8cm]{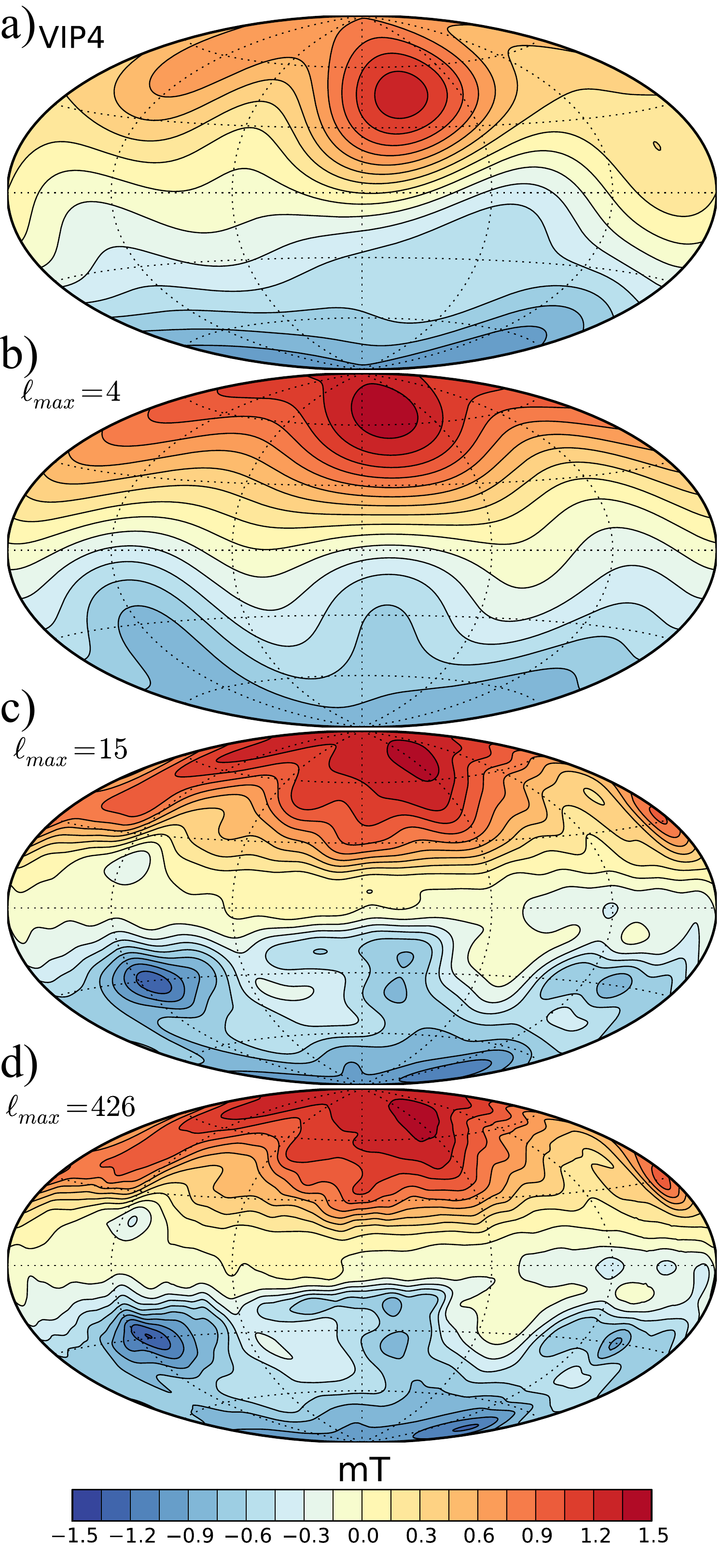}
 \caption{\scriptsize Comparison of the radial surface field of the Jupiter 
field model
VIP4 \citep{Connerney98} in panel a) with a selected snapshot from the
numerical simulation shown at three different spherical harmonic resolutions.
Panel b) depicts the same maximum spherical harmonic resolution $\ell_{max}=4$
as VIP4, panel c) illustrates the field at $\ell_{max}=15$, close to the
expected Juno detection limit,  while panel d) shows the full numerical solution
with $\ell_{max}=426$.}
 \label{fig:surffield}
\end{figure}

Fig.~\ref{fig:spectrum} and Fig.~\ref{fig:surffield} compare the magnetic field
in our simulation with models of Jupiter's magnetic field.
The Jovian magnetic field has been measured by several space missions but
since data taken below the magnetospheric standoff distance remain scarce
only the large scale field can be constrained. Observations of Io's auroral
footprint provide some additional information \citep{Connerney98,Hess11}.
Global magnetic fields are commonly decomposed into spherical harmonic
contributions with the Gauss coefficients $g_{\ell m}$ and $h_{\ell m}$
describing the internal magnetic field of spherical harmonic degree $\ell$ and
order $m$ at planetary surface \citep{Chapman40}.
The magnetic energy $W_{\ell m}$ carried by the pair $g_{\ell m}$ and
$h_{\ell m}$ then provides the energy for a specific degree and order:

\begin{equation}
\label{Blm}
W_{\ell m}=(\ell+1)\left( g_{\ell m}^2 + h_{\ell m}^2 \right).
\end{equation}
The energy spectrum per spherical harmonic degree or order can then
be calculated by summing the respective contributions:
\begin{equation}
\label{Bl}
W_{\ell}=\sum_{m=0}^{m=\ell} \,(\ell+1)\;\left( g_{\ell m}^2 + h_{\ell m}^2\right),
\end{equation}
\begin{equation}
\label{Bm}
W_{m}=\sum_{\ell=m}^{\ell=L} \,(\ell+1)\;\left( g_{\ell m}^2 + h_{\ell m}^2\right),
\end{equation}
where $L$ is the truncation degree of the model.
Fig.~\ref{fig:spectrum}a and b compare the magnetic energy
spectra of three Jupiter field models
with the time averaged spectra from our numerical simulation.
The Jupiter models VIP4 and VIT4 cover contributions up to $\ell=4$ while
the Ridley model reaches up to $\ell=7$.
The disagreement between the different models indicates that the data
are not sufficient to uniquely constrain degrees beyond $\ell=2$.
The regularization strongly affects degree $\ell=4$ contributions in VIP4 and VIT4
and $\ell=5$ and $\ell=6$ in the Ridley model.
The rescaled numerical simulation predicts a mean magnetic field strength that 
is about $40$\% stronger than the observation, a very reasonable agreement 
considering the model limitations.
The shape of the spectra agrees nicely with the field models.

%
To estimate the secular variation, we calculate the time
averaged spectral variation per year:
\begin{equation}
 \left<\dot{w}_\ell\right> =\left< |W_\ell(t+\delta t) - W_\ell(t)| / \delta t \right>
\end{equation}
where the triangular brackets denote the time average.
The respective spectrum is shown in
Fig.~\ref{fig:spectrum}c.
The variation amounts to about $0.1$\% at $\ell=1$ and
reaches a level of $0.01$\% of the dipole energy at $\ell=10$.
The high precision level of the Juno magnetometer should
allow to detect such a secular variation signal in at least the
small $\ell$ contributions (J.~Connerney, private communication).
Note however that the noise level shown in Fig.~\ref{fig:spectrum}c
assumes data compiled over the nominal one year mission duration.
The true secular variation noise level will thus be higher.

Fig.~\ref{fig:surffield} compares the magnetic field for a selected snapshot of
the simulation with the VIP4 internal field model \citep{Connerney98}.
The overall field structure up to spherical harmonic degree $\ell=4$
is very well captured by the numerical model. The dipole tilt for the
snapshot ($\Theta=6^\circ$) is somewhat lower than that for Jupiter
($\Theta=10.1^\circ$) but changes constantly over the simulation
with an average rate of $0.02$ degree per year.
The mean tilt is $\Theta=7.5^\circ$ and it reaches a
maximum of $\Theta=18.5^\circ$.
The rms surface field strength also
varies significantly around the mean value of $0.39\,$mT and
can double over a time span of about $500\,$yr. The mean
rate of change is somewhat slower at $0.1$\% per year.
Both the variation in tilt and field strength
are consistent with observations \citep{Russell01,Ridley12}.

A comparison of the radial magnetic field at the transition radius
(Fig.~\ref{fig:snapshot}c) and at the surface of the simulation
(Fig.~\ref{fig:surffield}d)
demonstrates that the higher harmonics decrease rapidly over the
weakly conducting outer envelope. However, the azimuthally extended
bands remain a clearly identifiable surface manifestation of the deeper dynamo
action associated to the equatorial jet.

A primary goal of the Juno mission is to significantly increase the knowledge of
the Jovian magnetic field. The spacecraft is scheduled to orbit Jupiter 30
times during its one year nominal mission duration. The first 15 orbits are
separated by $24^\circ$ in longitude so that a global but somewhat coarse
coverage is achieved after half the mission duration. The following 15 orbits
are shifted by $12^\circ$ in longitude to the first set.
The noise estimate shown in Fig.~\ref{fig:spectrum} assumes that
the magnetometer delivers a vector accuracy of $0.05$\% which is in
part limited by attitude information from the spacecraft's star camera.
However, the unique co-location of magnetometer and star camera at the end of
one of the three solar panels may allow an even higher accuracy.
The noise estimate furthermore assumes that external field contributions
can be modelled to an equivalent error level and that data points separated
by one minute flight time have uncorrelated errors and can thus be used as
independent data. The error propagation into the field model is calculated
for a standard least square fit of the data to the spherical harmonic
representation via Gauss coefficients (J.~Connerney, private
communication). This predicts that Juno data will
constrain the field up to $\ell=18$ or better . Fig.~\ref{fig:surffield}
demonstrates that for our simulation basically all the magnetic field features
are already captured at degree $15$. Furthermore magnetic banding due to the
equatorial jet is at times more pronounced than shown in
Fig.~\ref{fig:snapshot}. Thus, our results indicate that the Juno mission will
allow detection of the low latitude magnetic bands.

In conclusion, we present a numerical model for interior dynamics
that incorporates up-to-date knowledge of Jupiter's interior and is the
first to successfully reproduce the Jovian large scale magnetic field.
Previous attempts to model Jupiter's interior dynamics have indeed largely
failed to reproduce important features of the planet's large scale magnetic
field. They were either too simplistic with a too strong dipolar component
\citep{Heimpel11,Duarte13} or they were restricted to the outer envelope
dynamics producing a too axisymmetric and too little dipolar magnetic field
\citep{Stanley09}. Here, the combination of a deep-seated
dipolar dynamo
and a magnetic banding associated with the equatorial jet is
a key feature that distinguishes our model from these previous
numerical attempts.
The predictions of the magnetic field morphology at global and regional scales,
and of the secular variation, will allow our model to be tested against
Juno measurements.

\begin{acknowledgments}
We thank J.~Connerney for providing Juno's
magnetometer specifications and T.~Dannert for his development of the MPI
version of the MagIC code. All the computations have been carried out on the IBM
iDataPlex HPC System Hydra at the MPG Rechenzentrum Garching. TG and LD are
supported
by the Special Priority Program 1488 PlanetMag of the German Science
Foundation. MH is supported in part by an NSERC Discovery Grant.
\end{acknowledgments}

\bibliographystyle{agufull08}

%
%
\end{article}


\end{document}